\documentclass[runningheads]{llncs}

\usepackage[T1]{fontenc}
\usepackage{graphicx}
\usepackage{amsmath}
\usepackage{amssymb}
\usepackage{booktabs}
\usepackage{algorithm}
\usepackage{algorithmic}
\usepackage{multirow}
\usepackage{url}
\usepackage{pifont}
\usepackage[table]{xcolor}
\usepackage{etoolbox}

\newcommand{\cmark}{\ding{51}}
\newcommand{\xmark}{\ding{55}}

\definecolor{val_blue_lo}{RGB}{200, 230, 255}
\definecolor{val_blue_hi}{RGB}{130, 180, 255}
\definecolor{val_orange_lo}{RGB}{255, 230, 200}
\definecolor{val_orange_hi}{RGB}{255, 180, 120}
\definecolor{val_purple_lo}{RGB}{240, 210, 255}
\definecolor{val_purple_hi}{RGB}{200, 150, 240}

\newcommand{\blo}[1]{\cellcolor{val_blue_lo}\textbf{#1}}
\newcommand{\bhi}[1]{\cellcolor{val_blue_hi}\textbf{#1}}
\newcommand{\olo}[1]{\cellcolor{val_orange_lo}\textbf{#1}}
\newcommand{\ohi}[1]{\cellcolor{val_orange_hi}\textbf{#1}}
\newcommand{\plo}[1]{\cellcolor{val_purple_lo}\textbf{#1}}
\newcommand{\phii}[1]{\cellcolor{val_purple_hi}\textbf{#1}}

\begin{document}

\title{A Framework for Human-AI Q-Matrix Refinement: A NeuralCDM Evaluation}
\titlerunning{NeuralCDM Evaluation of LLM-Augmented Q-Matrices}

\author{Ying Zhang\inst{1} \and
Ningxi Cheng\inst{1} \and
Yizhu Gao\inst{1} \and
Hongmei Li\inst{1} \and
Lehong Shi\inst{1} \and
Nicholas Young\inst{1} \and
Geng Yuan\inst{1} \and
Xiaoming Zhai\inst{1}\thanks{Corresponding author.}}

\authorrunning{Y. Zhang et al.}

\institute{AI4STEM Education Center, University of Georgia, Athens, GA, USA\\
\email{\{zhying, EdwardCheng, yizhu.gao, hongmei.li1, ls77437, nicholas.young, geng.yuan, Xiaoming.Zhai\}@uga.edu}
}

\maketitle

\begin{abstract}
Q-matrices are a cornerstone of theory-driven assessment and learning analytics, because they make item demands--and students' underlying knowledge components and misconceptions--explicit and actionable. However, Q-matrices are typically crafted by experts, making them time-consuming to build, prone to subjectivity and rater disagreement, and difficult to validate empirically. We propose a framework for \emph{human-AI Q-matrix refinement} in which large language models (LLMs) function as a decision-support partner to generate candidate Q-matrices using structured, misconception-aware prompting, and NeuralCDM provides an empirical evaluation layer to compare candidates based on how well they explain student response data. As an illustrative case, we apply the framework to a thermodynamics assessment dataset and evaluate candidate Q-matrices by their ability to account for students' item-response patterns. Motivated by practical data-security constraints, we benchmark locally deployed LLMs against cloud-served models. This work positions LLMs as decision-support tools that surface plausible missing or misaligned skills/misconceptions, while empirical model fit serves as a guardrail against hallucinated structures.

\end{abstract}

\section{Introduction}
Q-Matrix serves as the foundational architecture in modern educational measurement and psychometrics. It is a binary mapping that specifies the latent mapping between assessment items and underlying Knowledge Components (KCs) or misconceptions \cite{de2011generalized,tatsuoka1983rule}. By explicitly defining the cognitive requirements of each item, the Q-Matrix enables Cognitive Diagnosis Models (CDMs) to transform raw student response data into detailed, multidimensional student profiles. This critical shift allows educators to move beyond simple total scores and understand exactly why a student succeeded or failed. The accuracy of student proficiency estimation depends critically on the quality of the Q-Matrix mapping \cite{rupp2008effects}. 

Despite its foundational importance, Q-Matrix construction remains a persistent challenge in the field \cite{chiu2013statistical}. The predominant approach relies on domain experts to manually specify item-level misconceptions and skills. This results in an inherently subjective and labor-intensive process. Prior research has documented several limitations of expert-driven Q-Matrix construction.

Firstly, experts often exhibit an expert blind spot due to their highly compiled knowledge structures, leading them to overlook common student misconceptions or implicit prerequisite skills. Secondly, inter-rater reliability in Q-Matrix labeling is often low, with different experts producing divergent specifications for the same items. Thirdly, the expert-driven construction of Q-matrices fails to scale to rapidly expanding item banks, thereby hindering the widespread deployment of cognitive diagnosis in large-scale online education \cite{desmarais2013matrix}.

To mitigate these challenges, data-driven approaches have been proposed, which implement statistical techniques to automatically recover or refine Q-Matrix entries from observed response patterns \cite{chen2015statistical}. Early works employed statistical estimation techniques that minimized the residual sum of squares \cite{chiu2013statistical}. Subsequently, matrix factorization methods were introduced to infer latent skill-item associations directly from response data \cite{sun2014alternating}. Additionally, statistical techniques such as the General Discrimination Index (GDI) was used to objectively validate attribute specifications \cite{de2016general}. More recently, machine learning and deep learning methods have been integrated into Neural Cognitive Diagnosis Models (NeuralCDM), which have demonstrated the capability to optimize Q-matrices as learnable parameters in an end-to-end fashion \cite{wang2020neural,gao2021rcd}. On the other hand, while such methods offer objectivity and scalability, they often lack interpretability and pedagogical transparency. An algorithm may flag a potential misspecification without explaining the underlying pedagogical reason or the specific misconception driving student errors. A statistical algorithm may optimize a mapping to improve model fit, but it cannot explain the specific pedagogical reason or the misconception driving student errors. This gap between statistical rigor and semantic transparency remains an open problem in CDM research, particularly in complex STEM domains like Thermodynamics, where student misconceptions are often deeply rooted and non-linearly related to item mastery.

The rapid advancement of Large Language Models (LLMs) offers a promising avenue for automating Q-Matrix construction and refinement. LLMs have exhibited profound capabilities in educational contexts, ranging from automated question generation to pedagogical content analysis \cite{latif2024knowledge}. Crucially, Chain-of-Thought (CoT) prompting facilitates interpretable problem solving, allowing models to simulate the expert tagging process \cite{lee2024applying,diao2024active}. This positions LLMs as semantic hypothesis generators, which are capable of inferring latent KCs attributes and anticipating student misconceptions directly through the semantic analysis of item content. However, LLMs outputs can be unreliable (e.g., hallucinations), so they cannot be treated as ground truth without empirical validation. This raises a fundamental challenge: how to leverage LLMs' semantic reasoning while maintaining the empirical validity demanded by psychometric applications. There is a critical need for an objective quantifier that can validate these AI-generated semantic hypotheses against actual student performance data.

To guide our investigation, we formulate the following research questions: 

\begin{itemize}
    \item RQ1: To what extent can LLMs generate a Q-Matrix that aligns with expert-defined item-misconception mappings for a thermodynamics concept inventory?

    \item RQ2: How do different prompting strategies (e.g., zero-shot vs. chain-of-thought reasoning) affect the alignment between LLM-generated and expert-defined Q-matrices? Does incorporating expert feedback into an iterative refinement process further improve the quality of LLM-generated Q-matrices beyond prompting strategies alone? 
    
    \item RQ3: Do the refined Q-matrices achieve better predictive performance in Cognitive Diagnosis Model compared to the original expert-defined Q-Matrix when evaluated against real student response data? 

\end{itemize}

\begin{figure}[t]
  \centering
  \includegraphics[width=0.8\linewidth]{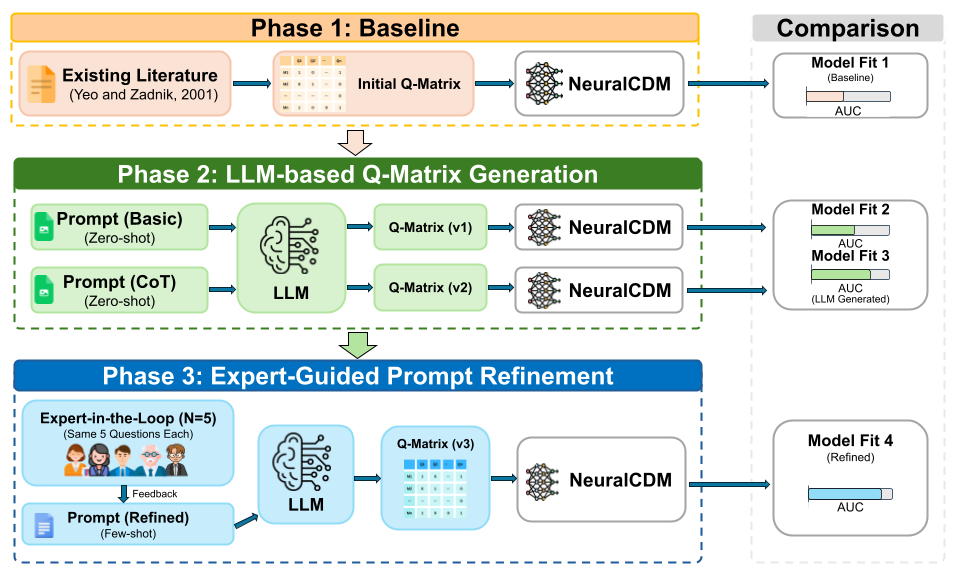}
  \caption{LLM-Augmented Q-Matrix Validation Pipeline}
  \label{fig:flowchart}
\end{figure}

In this paper, we propose a \emph{Generate-and-Validate} framework that addresses this question by combining LLM-based Q-Matrix generation with NeuralCDM validation \cite{wang2020neural}. Our approach transitions Q-Matrix construction from a manual labeling task to a scalable pipeline (As shown in Fig.~\ref{fig:flowchart}). 

To rigorously evaluate this framework, we implemented a dual-deployment experimental design that tests diverse prompting strategies (e.g., Chain-of-Thought, misconception-aware reasoning, and Expert-in-the-loop) across two distinct model classes. We utilize proprietary cloud-hosted models (e.g., GPT-4o, GPT-5) to establish an empirical upper bound for semantic reasoning, while simultaneously evaluating locally deployable open-weight models (e.g., Qwen3~\cite{qwen3} and Llama3~\cite{llama3}) as privacy-preserving alternatives. This comparative approach allows us to determine if cost-effective, offline-first models can achieve diagnostic parity with state-of-the-art cloud APIs, thereby ensuring that cognitive diagnosis remains accessible to institutions with strict data governance and budget constraints.

Subsequently, we employ NeuralCDM as an empirical comparator that evaluates candidate Q-matrices based on their alignment with real student response patterns. By modeling the non-linear student-item interactions, NeuralCDM empirically evaluates these candidates to identify the one that best fits the observed student response patterns. 

\section{Related Work}
\label{sec:RelatedWork}

\subsection{Construction of the Q-Matrix} 
\subsubsection{The Role of Q-Matrix in Diagnostic Assessment}

The Q-Matrix is the theoretical cornerstone of CDMs. It functions as a binary formalization ($\mathbf{Q} \in \{0, 1\}^{I \times K}$) that bridges $I$ assessment items with $K$ latent knowledge components (KCs) or misconceptions \cite{tatsuoka1983rule,de2011generalized}. In Intelligent Tutoring Systems (ITS), the Q-Matrix is indispensable for tracing learner mastery. For example, the domain of thermodynamics poses persistent conceptual challenges for learners. Common misconceptions include: confusion between heat and temperature; the belief that heat is a substance rather than energy transfer; the notion that temperature is an intrinsic property of materials (e.g., ``metals are naturally colder''); and misunderstandings about thermal equilibrium \cite{chi2005commonsense}. 

The most traditional approach to Q-Matrix development relies on domain experts to perform fine-grained task analysis. Experts decompose items into requisite skills based on pedagogical theories \cite{de2011generalized}. However, this approach is limited by subjective bias. Experts may hold idealized domain knowledge and overlook novice difficulties, and manual labeling can also exhibit low inter-rater reliability. As educational datasets expand, the labor-intensive nature of manual labeling also renders it unscalable for large-scale online assessment banks \cite{desmarais2013matrix}.

\subsubsection{Statistical and Data-Driven Refinement}
Over the decades, researchers have developed various methodologies to construct and refine the Q-Matrix, transitioning from manual expert annotation to automated, data-driven recovery. On one hand, to address the subjectivity of human labeling, researchers proposed statistical techniques to refine the Q-Matrix using student response data (e.g., Residual Sum of Squares \cite{chiu2013statistical}). On the other hand, data-driven methods can provide objectivity, but they may also reduce semantic interpretability, as an algorithm may flag an entry as mismatched without explaining the pedagogical reason (e.g., the misconception driving student errors).

Existing psychometric Q-Matrix validation methods also emerged in parallel. De la Torre and Chiu developed the General Discrimination Index (GDI) \cite{de2016general}, which provides a model-agnostic measure for empirically evaluating and validating individual Q-Matrix entries. Non-parametric classification methods have also been proposed to validate the attribute specifications without requiring a fully specified CDM \cite{chiu2013statistical}. While these approaches offer theoretically grounded validation within specific psychometric frameworks, they typically assume particular CDM structures (e.g., DINA) and may not fully capture the complex, non-linear interactions in STEM domains. In contrast, NeuralCDM offers flexibility in modeling student-item interactions through neural networks, which enables the direct comparison of candidate Q-Matrix via predictive fit metrics without restrictive structural assumptions. 

\subsection{Large Language Models for Cognitive Modeling}
Recent surveys highlight the efficacy of LLMs across diverse tasks, ranging from automated distractor generation to quality assurance for multiple-choice assessments \cite{kasneci2023chatgpt,wang2025autoscore}. The application of LLMs to cognitive modeling (particularly Q-Matrix construction) remains an emerging area with limited prior work. One study \cite{wang2025leveraging} employed GPT-4o-mini with a simulated textbook prompting strategy to extract KCs from multiple-choice questions, but their LLM-generated KC model achieved higher RMSE compared to expert-designed models, highlighting the challenges of fully automated 
KC extraction. Additionally, Chen et al.~\cite{chen2025llm} proposed LLM-CDM, which leverages prompt engineering to explore latent knowledge concepts from exercise texts, demonstrating the potential of LLMs to enrich semantic information in cognitive diagnosis. While these models demonstrate moderate alignment with expert-labeled ground truth, they often exhibit inconsistencies in handling domain-specific nuances, particularly when distinguishing between closely related concepts.

To enhance the interpretability and reliability of these annotations, Chain-of-Thought (CoT) prompting \cite{wei2022chain} has been adopted to elicit structured pedagogical reasoning. By forcing the model to articulate the step-by-step resolution of a problem before assigning a label, CoT aims to mimic the cognitive process of a domain expert.

However, a critical methodological gap persists in the literature: existing validations of LLM-generated structures rely predominantly on human-centric metrics, such as inter-rater agreement with human experts (e.g., Cohen's Kappa). Few studies have moved beyond expert consensus to validate LLM hypotheses against empirical student response data within a psychometric framework. Our work addresses this limitation by positioning the NeuralCDM \cite{wang2020neural} not merely as a diagnostic tool, but as an objective validator that grounds LLM-generated misconceptions in observed learning behaviors.

\subsection{Cognitive Diagnosis Models}

Classical CDMs such as the Deterministic Input, Noisy ``And'' Gate (DINA) \cite{de2009dina} assume simple conjunctive attribute-response relationships. In the DINA framework, a student is assumed to master an item only if they have mastered every latent attribute or knowledge component specified in the Q-Matrix for that item. This model accounts for deviations from this deterministic logic through two parameters: slipping (the probability that a student masters all required skills but still makes an error) and guessing (the probability that a student who lacks at least one required skill still answers correctly). However, it often fails to capture the complex, non-linear skill interactions prevalent in STEM domains because of its rigid conjunctive logic. 

To address this, Wang et al. \cite{wang2020neural} proposed NeuralCDM, which models student-item interactions through neural networks while strictly maintaining interpretability via monotonicity constraints. Although recent extensions like RCD \cite{gao2021rcd} and KaNCD \cite{wang2022neuralcd} have introduced relation maps and knowledge associations to enhance performance, they inevitably increase model complexity and rely heavily on high quality multi-dimensional data.

Consequently, NeuralCDM remains the most robust and generalizable foundation for our work, as it effectively balances deep representation learning with pedagogical interpretability without imposing restrictive data requirements. The flexible neural architecture also makes it the ideal candidate for integrating with CoT prompting, as it allows us to bridge the gap between continuous neural representations and structured pedagogical reasoning steps.

\subsection{Human-AI Collaboration} 
While LLMs possess strong semantic reasoning capabilities, they remain susceptible to intrinsic hallucination issues \cite{ji2023survey}. To mitigate this, the current studies emphasize Human-in-the-Loop frameworks, aiming to enhance the generative scale of AI with the rigor of human experts. The traditional paradigm of human-AI collaboration in education typically positions the expert as a post-hoc validator. For example, in automated item generation (AIG) and distractor generation tasks, LLMs are used to produce candidate content, which is subsequently filtered or edited by domain experts to ensure pedagogical validity \cite{lee2024applying}. While this approach improves quality, it retains a linear dependency on human effort: the expert's workload increases linearly with the size of the dataset. Such ``generate-then-edit'' workflows struggle to scale to large Q-matrices where thousands of item-attribute links require verification.

To address the scalability bottleneck, more recent Human-Computer Interaction (HCI) research proposes shifting expert intervention upstream, from correcting individual outputs to refining the model's instructions (prompts). Studies in interactive NLP demonstrate that users can effectively refine the model behavior by providing feedback on reasoning errors, which are then incorporated into refined prompts or few-shot examples. In the context of complex reasoning tasks, this iterative process enables experts to articulate robust knowledge that may not be captured by zero-shot instructions. For instance, recent work in legal and medical domains suggests that expert feedback on a small sample of CoT (Chain-of-Thought) reasoning paths can significantly calibrate the model's alignment with professional standards. However, the application of such expert-guided prompt calibration specifically for psychometric structure discovery (Q-Matrix construction) remains underexplored. To mitigate these risks without incurring the prohibitive costs of full manual annotation, we implemented a \emph{Lightweight Expert-in-the-Loop} protocol as shown in Fig~\ref{fig:flowchart}. This approach shifts the role of the expert from a high-volume annotator to an LLM proctor.

\section{Methodology}

\subsection{Dataset}
\subsubsection{Student Response}
We use thermodynamics assessment data with $N \approx 614$ student responses.
\subsubsection{Thermodynamic Q-Matrix}
To systematically assess the misconceptions in Thermodynamics, Yeo and Zadnik \cite{yeo2001introductory} developed the Thermal Concept Evaluation (TCE), a 26-item multiple-choice instrument targeting high school and introductory university students (approx.\ ages 15 to 19). Each item is situated in everyday contexts, with distractors corresponding to specific documented misconceptions. The TCE has demonstrated acceptable reliability (split-half $r = 0.81$) and discriminant validity across year levels. The complete item-to-misconception mapping is adopted as our expert-defined Q-Matrix (refer to \cite{yeo2001introductory}, Table I).

In this study, we adopt the TCE as our assessment instrument. The explicit mapping between distractors and misconceptions documented in the original TCE development provides a reference point for evaluating LLM-generated hypotheses. Following prior work \cite{tatsuoka1983rule}, we model misconceptions as binary diagnostic attributes alongside knowledge components, acknowledging this as a simplification that enables inference within the CDM framework.

\subsection{LLM as a Cognitive Annotator: Prompt Experiments}

To ensure the cognitive validity of the automated misconception analysis, we employed a rigorous, iterative prompt engineering strategy. This process evolved through four distinct developmental stages, transitioning from an open-ended inquiry to a highly constrained, expert-calibrated system.

\subsubsection{Phase 0: Preparation}
In the initial phase, we utilized a zero-shot approach, directly prompting the model to identify potential physics misconceptions without providing a specific taxonomy or confidence constraints. Qualitative evaluation revealed that while the model could identify general topic-related errors, it suffered from severe hallucinations and lack of precision. It failed to distinguish between plausible student errors and random guesses, rendering the output practically unusable for diagnostic purposes.

\subsubsection{Phase 1: Baseline}
To address the lack of precision, the next version (V1) introduced 1) a structured \textbf{Misconception Library} (Ids A1-D35) borrowed from \cite{yeo2001introductory} and 2) a \textbf{Confidence Calibration} mechanism. The model was instructed to rank misconceptions into High, Medium, and Low confidence tiers. While this version V1 improved output organization, it exhibited over-tagging behavior. The model tended to flag misconceptions simply because they shared keywords with the question stem (e.g., flagging Heat misconceptions whenever the word temperature appeared), without verifying if the specific cognitive trap was actually triggered.

\subsubsection{Phase 2: LLM Generation}
To curb this over-tagging, Version 2 (V2) introduced Critical Evaluation Principles as a logic filter. Unlike previous versions, V2 enforced strict negative constraints, requiring the model to validate each potential misconception against three criteria before labeling: (1) \textbf{Item-Content Alignment} (is the concept explicitly tested?), (2) \textbf{Error Attribution} (does it plausibly lead to a wrong answer?), and (3) \textbf{Distractor Mapping}. This phase shifted the model from pattern recognition to logical validation.

\subsubsection{Phase 3: Expert-Guided Prompt Refinement}
Finally, the optimized Version 3 (V3) aimed to bridge the gap between rule-following and expert intuition. We engaged five physics education researchers, each holding a doctoral degree, to annotate a subset of items. Each of them annotated the same subset of five items (25 response options) together with their reasoning, which were later added into Few-Shot Examples. Their annotations were synthesized through discussion into representative Few-Shot examples that demonstrated reasoning to the model. Table \ref{tab:prompt_evolution} summarizes this evolutionary process.

\begin{table}[ht]
    \centering
    \caption{Evolution of Prompt Engineering Strategy: Feature Comparison across Versions}
    \label{tab:prompt_evolution}
    \footnotesize
    \setlength{\tabcolsep}{4pt}
    
    \begin{tabular}{@{}lcccc@{}}
        \toprule
        \textbf{Features} & \textbf{V0} & \textbf{V1} & \textbf{V2} & \textbf{V3} \\
        & \scriptsize(Baseline) & \scriptsize(Structured) & \scriptsize(Rule-Based) & \scriptsize(Expert-FewShot) \\
        \midrule
        Misconception Library & \xmark & \cmark & \cmark & \cmark \\
        Confidence Calibration & \xmark & \cmark & \cmark & \cmark \\
        Evaluation Principles & \xmark & \xmark & \cmark & \cmark \\
        Few-Shot Examples & \xmark & \xmark & \xmark & \cmark \\
        \midrule
        Reasoning Depth & Shallow & Structured & Logical & Expert-Sim. \\
        \bottomrule
    \end{tabular}
\end{table}

\subsection{NeuralCDM Model Fit}
Wang et al. (2020) proposed NeuralCDM, which replaces the rigid linear assumptions of traditional CDMs (e.g., DINA) with multi-layer neural networks while maintaining interpretability through monotonicity constraints \cite{wang2020neural}. A feature of NeuralCDM is its ability to treat the Q-Matrix as a learnable parameter, facilitating end-to-end optimization against massive student response datasets \cite{gao2021rcd}. 

Unlike traditional methods that focus solely on parameter estimation, deep learning algorithms can be used as a validation guardrail. For instance, model fit metrics (e.g., AUC, and RMSE) derived from NeuralCDM can be used to compare candidate Q-matrices. The model fit is a measure of how effectively a model (e.g., Q-Matrix) explains the observed patterns in the actual student data. For instance, the model fit metrics AUC measures the model's ability to distinguish between correct and incorrect responses; RMSE tracks the error between predicted and actual scores. These metrics quantify the alignment between the cognitive hypotheses and real-world outcomes. If a refined Q-Matrix achieves higher predictive accuracy on the student data, it provides empirical evidence that the proposed cognitive structure is more consistent with actual learner behaviors. This positions NeuralCDM not only as a diagnostic tool but as a rigorous empirical quantifier for evaluating AI-generated or expert-refined Q-Matrix hypotheses.

In our study, we used NeuralCDM as an empirical comparator. Importantly, we do not treat better model fit as definitive evidence of cognitive validity, but rather as one source of empirical support indicating greater consistency with student performance data.

\subsection{Dual-Deployment Framework: Cloud vs. Local Inference}

While cloud-based Large Language Models (e.g., GPT-5) currently define the state-of-the-art in semantic reasoning, their direct integration into educational infrastructures faces significant barriers.
First, strict data governance regulations (e.g., GDPR \cite{gdpr}, FERPA) preclude the transmission of sensitive student response data or proprietary assessment items to third-party API endpoints.
Second, the operational costs of commercial APIs scale linearly with the volume of assessment items and the complexity of Chain-of-Thought prompting. This creates a financial bottleneck that may render continuous, large-scale cognitive diagnosis unsustainable for under-resourced institutions.

To address these constraints, we engineered our \emph{Generate-and-Validate} pipeline to support a dual-deployment strategy, seamlessly interfacing with both cloud-hosted APIs and locally deployed open-source models (e.g., Llama-3 \cite{llama3}, Qwen3 \cite{qwen3}).
This architecture is essential for our comparative analysis. 
We utilize high-parameter cloud models to establish an empirical ``upper bound'' for reasoning capability and Q-Matrix quality. 
Simultaneously, we deploy local models to evaluate their efficacy as a privacy-preserving alternative.
By benchmarking local models against this cloud-based ceiling, we aim to validate a cost-effective, offline-first solution that ensures data sovereignty without compromising the psychometric validity and pedagogical interpretability of the generated Q-Matrix structure.

\begin{table}[ht]
\centering
\caption{Q-Matrix Comparison Stats. \textbf{Colored cells} indicate the best value in that column for Local and Cloud groups respectively. Lighter shades denote Local models; Darker shades denote Cloud models.}
\label{tab:qmatrix_stats}

\resizebox{\textwidth}{!}{%
\begin{tabular}{l l c c c c c c c c c c}
\toprule
\textbf{Model} & \textbf{Config} & \textbf{TP} & \textbf{FP} & \textbf{FN} & \textbf{TN} & \textbf{TPR} & \textbf{TNR} & \textbf{Precision} & \textbf{Recall} & \textbf{Micro F1} & \textbf{Macro F1} \\
\midrule

Llama-3.1-8B & High-only & 14 & \blo{43} & 62 & \blo{765} & 0.1842 & \blo{0.9468} & 0.2456 & 0.1842 & 0.2105 & 0.5801 \\
Qwen3-30B & High-only & \blo{27} & 70 & \blo{49} & 738 & \blo{0.3553} & 0.9134 & 0.2784 & \blo{0.3553} & \blo{0.3121} & \blo{0.6213} \\
Qwen3-8B & High-only & 24 & 58 & 52 & 750 & 0.3158 & 0.9282 & \blo{0.2927} & 0.3158 & 0.3038 & 0.6203 \\
\noalign{\smallskip}
gpt-4o & High-only & 23 & \bhi{31} & 53 & \bhi{777} & 0.3026 & \bhi{0.9616} & 0.4259 & 0.3026 & 0.3538 & 0.6582 \\
gpt-5 & High-only & \bhi{41} & 52 & \bhi{35} & 756 & \bhi{0.5395} & 0.9356 & 0.4409 & \bhi{0.5395} & \bhi{0.4852} & \bhi{0.746} \\
gpt-5.1 & High-only & 36 & 41 & 40 & 767 & 0.4737 & 0.9493 & \bhi{0.4675} & 0.4737 & 0.4706 & 0.7198 \\

\midrule

Llama-3.1-8B & High+Med & 18 & \olo{85} & 58 & \olo{723} & 0.2368 & \olo{0.8948} & 0.1748 & 0.2368 & 0.2011 & 0.5659 \\
Qwen3-30B & High+Med & \olo{37} & 151 & \olo{39} & 657 & \olo{0.4868} & 0.8131 & 0.1968 & \olo{0.4868} & \olo{0.2803} & 0.5797 \\
Qwen3-8B & High+Med & 32 & 124 & 44 & 684 & 0.4211 & 0.8465 & \olo{0.2051} & 0.4211 & 0.2759 & \olo{0.5861} \\
\noalign{\smallskip}
gpt-4o & High+Med & 31 & \ohi{69} & 45 & \ohi{739} & 0.4079 & \ohi{0.9146} & 0.31 & 0.4079 & 0.3523 & 0.6528 \\
gpt-5 & High+Med & \ohi{56} & 126 & \ohi{20} & 682 & \ohi{0.7368} & 0.8441 & 0.3077 & \ohi{0.7368} & 0.4341 & \ohi{0.7023} \\
gpt-5.1 & High+Med & 50 & 96 & 26 & 712 & 0.6579 & 0.8812 & \ohi{0.3425} & 0.6579 & \ohi{0.4505} & 0.6984 \\

\midrule

Llama-3.1-8B & All Tiers & 23 & \plo{130} & 53 & \plo{678} & 0.3026 & \plo{0.8391} & \plo{0.1503} & 0.3026 & 0.2009 & \plo{0.5587} \\
Qwen3-30B & All Tiers & \plo{41} & 235 & \plo{35} & 573 & \plo{0.5395} & 0.7092 & 0.1486 & \plo{0.5395} & \plo{0.233} & 0.5264 \\
Qwen3-8B & All Tiers & 32 & 184 & 44 & 624 & 0.4211 & 0.7723 & 0.1481 & 0.4211 & 0.2192 & 0.5449 \\
\noalign{\smallskip}
gpt-4o & All Tiers & 38 & \phii{109} & 38 & \phii{699} & 0.5 & \phii{0.8651} & 0.2585 & 0.5 & 0.3408 & 0.6355 \\
gpt-5 & All Tiers & \phii{61} & 188 & \phii{15} & 620 & \phii{0.8026} & 0.7673 & 0.245 & \phii{0.8026} & 0.3754 & 0.6459 \\
gpt-5.1 & All Tiers & 56 & 145 & 20 & 663 & 0.7368 & 0.8205 & \phii{0.2786} & 0.7368 & \phii{0.4043} & \phii{0.6604} \\

\bottomrule
\end{tabular}%
}
\end{table}

\section{Experiments and Results}
\subsection{Experimental Setup}

\subsubsection{Local Deployment Environment.}
Local inference was performed on a high-performance computing node equipped with 4 $\times$ NVIDIA L40S GPUs (48GB VRAM per GPU). We utilized vLLM as the inference engine \cite{vllm}, deployed as an OpenAI-compatible server. To ensure the syntactic validity of the generated Q-Matrices, we utilized structured output feature via the OpenAI-compatible API endpoint, which enforces context-free grammar constraints to guarantee valid JSON outputs.

\subsubsection{Model Selection.}
We evaluated a diverse set of models queried with a temperature of 0.0 to maximize determinism. For local open-weight inference, we selected Qwen3-30B-A3B-Instruct, Qwen3-8B, and Llama 3.1 8B. For proprietary cloud-hosted baselines, we utilized the official OpenAI API to access GPT-4o, GPT-5, and GPT-5.1.

\begin{table}[t]
\caption{NeuralCDM performance metrics. \textbf{Colored cells} indicate the best value in that column for Local and Cloud groups respectively. Lighter shades denote Local models; Darker shades denote Cloud models.}
\label{tab:comprehensive_results}
\centering
\scriptsize
\setlength{\tabcolsep}{1.5pt}

\begin{tabular}{lc cc cc cc}
\toprule
 & & \multicolumn{2}{c}{High Confidence} & \multicolumn{2}{c}{Medium/High} & \multicolumn{2}{c}{All Choices} \\
\cmidrule(lr){3-4} \cmidrule(lr){5-6} \cmidrule(lr){7-8}
Model & Prompt & AUC & RMSE & AUC & RMSE & AUC & RMSE \\
\midrule

\multicolumn{2}{l}{\textit{Expert Baseline}} & $0.717 \: \text{\tiny $\pm \mathit{.011}$}$ & $0.400 \: \text{\tiny $\pm \mathit{.002}$}$ & -- & -- & -- & -- \\
\multicolumn{2}{l}{\textit{GPT-5 (v0)}} & $0.642 \: \text{\tiny $\pm \mathit{.026}$}$ & $0.614 \: \text{\tiny $\pm \mathit{.025}$}$ & -- & -- & -- & -- \\

\midrule
\multicolumn{8}{l}{\textbf{Open-Weights (Local)}} \\ \addlinespace[2pt]
Qwen3-8B & v1 & $0.692 \: \text{\tiny $\pm \mathit{.014}$}$ & $0.576 \: \text{\tiny $\pm \mathit{.015}$}$ & \olo{$0.707 \: \text{\tiny $\pm \mathit{.020}$}$} & \olo{$0.533 \: \text{\tiny $\pm \mathit{.015}$}$} & \plo{$0.707 \: \text{\tiny $\pm \mathit{.032}$}$} & \plo{$0.515 \: \text{\tiny $\pm \mathit{.009}$}$} \\
 & v2 & $0.722 \: \text{\tiny $\pm \mathit{.005}$}$ & $0.404 \: \text{\tiny $\pm \mathit{.002}$}$ & $0.723 \: \text{\tiny $\pm \mathit{.011}$}$ & $0.406 \: \text{\tiny $\pm \mathit{.004}$}$ & $0.726 \: \text{\tiny $\pm \mathit{.009}$}$ & $0.406 \: \text{\tiny $\pm \mathit{.003}$}$ \\
 & v3 & \blo{$0.752 \: \text{\tiny $\pm \mathit{.005}$}$} & \blo{$0.396 \: \text{\tiny $\pm \mathit{.003}$}$} & \olo{$0.753 \: \text{\tiny $\pm \mathit{.011}$}$} & $0.403 \: \text{\tiny $\pm \mathit{.005}$}$ & $0.754 \: \text{\tiny $\pm \mathit{.009}$}$ & $0.406 \: \text{\tiny $\pm \mathit{.004}$}$ \\
\addlinespace[3pt]
Llama3.1-8B & v1 & $0.667 \: \text{\tiny $\pm \mathit{.013}$}$ & $0.587 \: \text{\tiny $\pm \mathit{.010}$}$ & $0.683 \: \text{\tiny $\pm \mathit{.025}$}$ & $0.560 \: \text{\tiny $\pm \mathit{.021}$}$ & $0.699 \: \text{\tiny $\pm \mathit{.017}$}$ & $0.544 \: \text{\tiny $\pm \mathit{.026}$}$ \\
 & v2 & $0.724 \: \text{\tiny $\pm \mathit{.003}$}$ & \blo{$0.394 \: \text{\tiny $\pm \mathit{.002}$}$} & $0.729 \: \text{\tiny $\pm \mathit{.005}$}$ & \olo{$0.400 \: \text{\tiny $\pm \mathit{.002}$}$} & $0.734 \: \text{\tiny $\pm \mathit{.008}$}$ & $0.402 \: \text{\tiny $\pm \mathit{.003}$}$ \\
 & v3 & $0.741 \: \text{\tiny $\pm \mathit{.008}$}$ & \blo{$0.396 \: \text{\tiny $\pm \mathit{.003}$}$} & $0.744 \: \text{\tiny $\pm \mathit{.009}$}$ & \olo{$0.400 \: \text{\tiny $\pm \mathit{.004}$}$} & $0.758 \: \text{\tiny $\pm \mathit{.005}$}$ & \plo{$0.399 \: \text{\tiny $\pm \mathit{.001}$}$} \\
\addlinespace[3pt]
Qwen3-30B & v1 & \blo{$0.705 \: \text{\tiny $\pm \mathit{.013}$}$} & \blo{$0.544 \: \text{\tiny $\pm \mathit{.013}$}$} & $0.685 \: \text{\tiny $\pm \mathit{.018}$}$ & $0.534 \: \text{\tiny $\pm \mathit{.013}$}$ & $0.682 \: \text{\tiny $\pm \mathit{.013}$}$ & $0.521 \: \text{\tiny $\pm \mathit{.010}$}$ \\
 & v2 & \blo{$0.733 \: \text{\tiny $\pm \mathit{.011}$}$} & $0.397 \: \text{\tiny $\pm \mathit{.004}$}$ & \olo{$0.743 \: \text{\tiny $\pm \mathit{.011}$}$} & $0.402 \: \text{\tiny $\pm \mathit{.004}$}$ & \plo{$0.750 \: \text{\tiny $\pm \mathit{.010}$}$} & \plo{$0.400 \: \text{\tiny $\pm \mathit{.006}$}$} \\
 & v3 & $0.737 \: \text{\tiny $\pm \mathit{.009}$}$ & $0.398 \: \text{\tiny $\pm \mathit{.002}$}$ & $0.748 \: \text{\tiny $\pm \mathit{.007}$}$ & $0.405 \: \text{\tiny $\pm \mathit{.005}$}$ & \plo{$0.762 \: \text{\tiny $\pm \mathit{.006}$}$} & \plo{$0.399 \: \text{\tiny $\pm \mathit{.004}$}$} \\
\addlinespace[3pt]

\midrule
\multicolumn{8}{l}{\textbf{Cloud APIs}} \\ \addlinespace[2pt]
GPT-4o & v1 & $0.668 \: \text{\tiny $\pm \mathit{.022}$}$ & $0.664 \: \text{\tiny $\pm \mathit{.027}$}$ & $0.703 \: \text{\tiny $\pm \mathit{.043}$}$ & $0.556 \: \text{\tiny $\pm \mathit{.029}$}$ & \phii{$0.720 \: \text{\tiny $\pm \mathit{.027}$}$} & $0.543 \: \text{\tiny $\pm \mathit{.024}$}$ \\
 & v2 & $0.704 \: \text{\tiny $\pm \mathit{.010}$}$ & \bhi{$0.394 \: \text{\tiny $\pm \mathit{.006}$}$} & $0.716 \: \text{\tiny $\pm \mathit{.008}$}$ & $0.406 \: \text{\tiny $\pm \mathit{.003}$}$ & $0.734 \: \text{\tiny $\pm \mathit{.010}$}$ & $0.413 \: \text{\tiny $\pm \mathit{.004}$}$ \\
 & v3 & $0.740 \: \text{\tiny $\pm \mathit{.006}$}$ & $0.395 \: \text{\tiny $\pm \mathit{.001}$}$ & $0.742 \: \text{\tiny $\pm \mathit{.011}$}$ & \ohi{$0.397 \: \text{\tiny $\pm \mathit{.003}$}$} & $0.738 \: \text{\tiny $\pm \mathit{.011}$}$ & $0.406 \: \text{\tiny $\pm \mathit{.003}$}$ \\
\addlinespace[3pt]
GPT-5 & v1 & \bhi{$0.715 \: \text{\tiny $\pm \mathit{.022}$}$} & \bhi{$0.549 \: \text{\tiny $\pm \mathit{.027}$}$} & $0.702 \: \text{\tiny $\pm \mathit{.013}$}$ & \ohi{$0.538 \: \text{\tiny $\pm \mathit{.012}$}$} & $0.701 \: \text{\tiny $\pm \mathit{.020}$}$ & \phii{$0.525 \: \text{\tiny $\pm \mathit{.013}$}$} \\
 & v2 & \bhi{$0.727 \: \text{\tiny $\pm \mathit{.009}$}$} & $0.398 \: \text{\tiny $\pm \mathit{.002}$}$ & \ohi{$0.743 \: \text{\tiny $\pm \mathit{.006}$}$} & \ohi{$0.402 \: \text{\tiny $\pm \mathit{.003}$}$} & \phii{$0.749 \: \text{\tiny $\pm \mathit{.007}$}$} & \phii{$0.406 \: \text{\tiny $\pm \mathit{.004}$}$} \\
 & v3 & \bhi{$0.780 \: \text{\tiny $\pm \mathit{.008}$}$} & \bhi{$0.386 \: \text{\tiny $\pm \mathit{.002}$}$} & \ohi{$0.776 \: \text{\tiny $\pm \mathit{.006}$}$} & \ohi{$0.397 \: \text{\tiny $\pm \mathit{.005}$}$} & $0.775 \: \text{\tiny $\pm \mathit{.007}$}$ & $0.405 \: \text{\tiny $\pm \mathit{.005}$}$ \\
\addlinespace[3pt]
GPT-5.1 & v1 & $0.673 \: \text{\tiny $\pm \mathit{.014}$}$ & $0.584 \: \text{\tiny $\pm \mathit{.025}$}$ & \ohi{$0.707 \: \text{\tiny $\pm \mathit{.025}$}$} & $0.542 \: \text{\tiny $\pm \mathit{.016}$}$ & $0.708 \: \text{\tiny $\pm \mathit{.024}$}$ & \phii{$0.525 \: \text{\tiny $\pm \mathit{.019}$}$} \\
 & v2 & $0.718 \: \text{\tiny $\pm \mathit{.006}$}$ & $0.400 \: \text{\tiny $\pm \mathit{.003}$}$ & $0.728 \: \text{\tiny $\pm \mathit{.005}$}$ & $0.410 \: \text{\tiny $\pm \mathit{.004}$}$ & $0.743 \: \text{\tiny $\pm \mathit{.009}$}$ & $0.411 \: \text{\tiny $\pm \mathit{.004}$}$ \\
 & v3 & $0.749 \: \text{\tiny $\pm \mathit{.005}$}$ & $0.393 \: \text{\tiny $\pm \mathit{.002}$}$ & $0.755 \: \text{\tiny $\pm \mathit{.010}$}$ & $0.404 \: \text{\tiny $\pm \mathit{.003}$}$ & \phii{$0.778 \: \text{\tiny $\pm \mathit{.007}$}$} & \phii{$0.402 \: \text{\tiny $\pm \mathit{.003}$}$} \\
\bottomrule
\end{tabular}
\end{table}

\subsection{Consistency Analysis Between Expert and LLM Matrices}

To evaluate the semantic alignment between the automated pipelines and human domain experts, we utilized standard confusion matrix metrics, treating the expert-defined Q-Matrix \cite{yeo2001introductory} as the reference standard. We calculated True Positives (TP), False Positives (FP), and False Negatives (FN) to derive Precision, Recall, and F1-scores across different model families and confidence tiers. These metrics quantify the extent to which LLMs can autonomously reproduce the misconception mappings identified by physics educators.

As detailed in Table~\ref{tab:qmatrix_stats}, there is a marked disparity in alignment between cloud-hosted and locally deployed models. The proprietary cloud models (e.g., GPT-5, GPT-4o) consistently achieved higher alignment with the expert baseline compared to open-weight local alternatives. For instance, in the \textit{High-Only} confidence configuration, GPT-5 achieved the highest Recall ($0.5395$) and Micro-F1 ($0.4852$), significantly outperforming the Llama-3.1-8B baseline (Recall=$0.1842$; Micro-F1=$0.2105$). This suggests that larger, cloud-based models possess a stronger semantic grasp of the specific thermodynamic concepts defined by experts, allowing them to replicate the existing coding scheme with greater fidelity.

It is critical, however, to interpret these consistency metrics with nuance. While high overlap (high Recall/F1) indicates that the model successfully mimics expert judgment, strictly penalizing deviations (low Precision or high False Positives) may be misleading in an exploratory context. A ``False Positive'' in this framework simply denotes a misconception flagged by the LLM that was absent in the original expert list; this does not necessarily constitute a hallucination. It may instead represent a valid latent knowledge component or a subtle misconception that experts overlooked or aggregated differently. Consequently, we frame these metrics as a measure of \textit{reproductive consistency} rather than absolute diagnostic quality.

\subsection{Comparative Analysis of Candidate Q-Matrices}
To assess the pedagogical validity of the generated Q-matrices, we employed NeuralCDM \cite{wang2020neural} as an empirical validator. We utilized two primary metrics to quantify model fit: \textit{Area Under the Curve (AUC)}, which measures the model's ability to distinguish between correct and incorrect student responses, and \textit{Root Mean Square Error (RMSE)}, which quantifies the divergence between predicted response probabilities and observed outcomes. A higher AUC and lower RMSE indicate that the Q-Matrix structure more accurately reflects the latent cognitive processes driving student performance.

\subsubsection{Cloud-Hosted Models and the ``High-Confidence'' Advantage.}
As presented in Table~\ref{tab:comprehensive_results}, the cloud-hosted models demonstrated superior performance when restricted to their highest confidence outputs. Specifically, the Q-Matrix generated by \textbf{GPT-5 (V3)} using only ``High Confidence'' labels achieved a global best AUC of \textbf{0.780} and an RMSE of \textbf{0.386}. This represents a substantial improvement over the original Expert Baseline (AUC=0.717; RMSE=0.400). This result suggests that when state-of-the-art models express high certainty, their semantic reasoning aligns exceptionally well with the empirical patterns of student misconceptions, effectively pruning weak or ambiguous signals that may exist in the full expert-defined matrix.

\subsubsection{Parameter Efficiency of Local Models.}
Remarkably, locally deployed open-weight models proved to be highly competitive alternatives despite being orders of magnitude smaller than the proprietary cloud baselines. For instance, \textbf{Qwen3-8B} (V3, High-Only) achieved an AUC of \textbf{0.752}, surpassing the expert baseline (AUC=0.717) while operating with a fraction of the parameter count of models like GPT-4o or GPT-5. This finding challenges the assumption that high-quality LLM-based Q-matrix generation requires the massive computational footprint of frontier models. It demonstrates that smaller, privacy-preserving models---when guided by optimized prompting (V3) and confidence filtering---can capture the core semantic structures of the domain as effectively as much larger architectures, making them a viable solution for resource-constrained educational environments.

\subsection{Role of Prompt Engineering}
Prompt engineering proved critical for transforming raw LLM capabilities into psychometrically valid outputs. As shown in Table~\ref{tab:comprehensive_results}, the progression from V1 to V3 yielded consistent improvements in NeuralCDM model fit across all models.

For cloud models, GPT-5's AUC improved from 0.715 (V1) to 0.780 (V3) in the High-confidence tier, a gain of 6.5\%. For local models, Qwen3-8B improved from 0.692 (V1) to 0.752 (V3), demonstrating that prompt optimization benefits both model classes. The V3 expert-guided strategy achieved the best results across nearly all configurations, confirming that few-shot examples from domain experts successfully calibrate LLM reasoning toward pedagogically valid misconception identification.

\section{Discussion}

Our core contribution is a practical framework for \emph{human--AI Q-matrix refinement} that connects three elements: (i) LLMs as semantic hypothesis generators that propose candidate item--attribute links, (ii) NeuralCDM as an empirical evaluation layer that compares candidates based on consistency with observed response patterns, and (iii) lightweight expert input used selectively to calibrate prompts rather than exhaustively label entries. This framing shifts Q-matrix development from a one-shot expert artifact to an iterative, evidence-informed workflow that can be repeated as item banks evolve.

\subsection{Human--AI Collaboration}
We frame the LLM as a decision-support tool rather than an autonomous annotator: it proposes refinements and the domain experts review disagreements. This can shift expert work from ``labeling from scratch'' to ``reviewing high-probability suggestions.''

The framework operationalizes human--AI collaboration through a clear division of labor. The LLM contributes breadth by rapidly generating candidate mappings and surfacing plausible missing or misaligned misconceptions. NeuralCDM contributes a data-grounded filter by adjudicating among candidates using predictive fit. Experts contribute depth by providing targeted feedback on a small subset of items to correct systematic reasoning errors (e.g., keyword-driven over-tagging), which then improves subsequent generations through prompt calibration. Importantly, expert effort does not scale linearly with the number of items; instead, it concentrates on high-leverage prompt failures and ambiguous cases. This efficiency gain is particularly relevant for large-scale online learning platforms where item banks may contain thousands of questions and traditional expert annotation becomes prohibitively expensive.

\subsection{Limitations and Threats to Validity}
This study has several limitations. First, the evaluation is conducted on a single thermodynamics instrument, and results may vary across domains with different misconception structures or item formats. Second, our ``expert baseline'' is treated as a reference point but may itself contain omissions or ambiguities; disagreement with experts does not necessarily indicate LLM error. Third, NeuralCDM-based comparison depends on modeling assumptions and the size/representativeness of the response dataset; fit differences may shrink or change under alternative CDMs or cross-validation schemes. Fourth, prompt variants and model families are not exhaustive; additional prompt controls (e.g., self-consistency, retrieval-augmented prompts using curriculum materials) may further affect outcomes.

\section{Conclusion}
We have presented a framework for human--AI collaboration in Q-matrix refinement. The approach combines three components: LLMs generate candidate item--attribute mappings through structured prompting, NeuralCDM evaluates candidates against observed student response data, and domain experts provide targeted feedback to calibrate prompts. This division of labor shifts Q-matrix construction from exhaustive manual labeling toward selective expert oversight of AI-generated hypotheses. Empirical evaluation on a thermodynamics concept inventory demonstrated that iterative prompt refinement (V1, V2, V3) improved both expert alignment and NeuralCDM model fit. The best-performing configurations exceeded the expert-baseline AUC (0.780 vs.\ 0.717), and locally deployed models achieved results comparable to cloud APIs, supporting privacy-preserving deployment. We emphasize that improved model fit does not establish cognitive validity; it indicates consistency with response data. The framework is best understood as a method for generating empirically grounded hypotheses that warrant further expert review. Future work should examine generalization across domains, integrate qualitative evaluation of generated labels, and assess downstream effects on diagnostic feedback and learning outcomes.

\begin{credits}
\subsubsection{\ackname}
This work was supported by the Institute of Education Sciences (GENIUS, \#R305C240010) and the National Science Foundation (TALENT, \#2507128). The opinions expressed are those of the authors and do not represent views of the funding agencies.

\subsubsection{\discintname}
The authors have no competing interests to declare that are relevant to the content of this article.
\end{credits}

\bibliographystyle{splncs04}
\bibliography{references}

\end{document}